\documentclass[prd,nofootinbib,showpacs,preprint]{revtex4}
\usepackage[T1]{fontenc}
\usepackage{amsmath,amssymb}
\usepackage{epsfig}
\usepackage{url}
\usepackage{graphicx}
\usepackage[usenames,dvipsnames]{color}
\usepackage{slashed}
\usepackage[colorlinks,citecolor=blue]{hyperref}
\usepackage{color}
\usepackage{cancel}

\usepackage{epstopdf} %converting to PDF
\usepackage{graphicx}
\usepackage[usenames,dvipsnames]{color}
\usepackage[a4paper,left=2.3cm,right=2.3cm]{geometry}
\usepackage{amsmath}
\usepackage{float}
\usepackage[latin1]{inputenc}
\usepackage{amsmath}
\usepackage{amsfonts}
\usepackage{amssymb}
\usepackage{epsfig}
\usepackage[toc,page]{appendix}
\newcommand{\be}{\begin{equation}}
\newcommand{\ee}{\end{equation}}
\newcommand{\bea}{\begin{eqnarray}}
\newcommand{\eea}{\end{eqnarray}}

\begin{document}
\preprint{HRI-RECAPP-2018-003}
\title{Implications of a vector-like lepton doublet and  scalar \\ Leptoquark on $R(D^{(*)})$.}
\author{Lobsang Dhargyal} 
\email{dhyargal@hri.res.in}
\affiliation{Regional Centre for Accelerator-based Particle Physics,\\
Harish-Chandra Research Institute, HBNI, Jhusi, Allahabad - 211019, India}
\author{Santosh Kumar Rai} 
\email{skrai@hri.res.in}
\affiliation{Regional Centre for Accelerator-based Particle Physics,\\
Harish-Chandra Research Institute, HBNI, Jhusi, Allahabad - 211019, India}
\date{\today}

\begin{abstract}

We study the phenomenological constraints and consequences in the flavor sector, 
of introducing a new fourth generation $\mathcal{Z}_{2}$ odd vector-like lepton doublet 
along with a Standard Model (SM) singlet scalar and an $SU(2)_{L}$ singlet scalar 
leptoquark carrying  electromagnetic charge of $+2/3$, both odd under a 
$\mathcal{Z}_{2}$. We show that with little fine tuning among the various Yukawa 
couplings in the new physics (NP) Lagrangian along with the CKM parameters, the model is 
able to push the theoretical value of $R(D^{*})^{th}$ from $0.252 \pm 0.003$ to 
$0.263 \pm 0.051$ and $R(D)^{th}$ from $0.300 \pm 0.008$ to $0.313 \pm 0.158$ 
compared to the SM value. Especially the NP contributions are able to reduce 
the discrepancy between experiment and theory of $R(D^{*})$ substantially compared 
to SM. This is quite impressive given that the model satisfy all other very stringent 
constrains coming from neutral meson oscillations and precision $Z$-pole data.

\end{abstract}
\maketitle

\section{Introduction.}

The Standard Model (SM) of particle physics has been very successful in accounting for 
particle interactions and gives an admissible explanation for the electroweak symmetry breaking (EWSB) mechanism that agrees with all observed data. It has also been tested 
to a very high degree of precision and the discovery of the Higgs boson at the Large Hadron
Collider (LHC) completed the last missing piece of the framework. 
However there are also several experimental observations such as that of 
neutrino-oscillation suggesting neutinos to have mass, existence of dark-matter (DM) 
and dark-energy in the Universe which definitely point to the incompleteness of our 
understanding and to physics beyond the SM (BSM). With very little hint of any new physics 
(NP) discovery by direct searches at LHC at the moment there is huge interest in 
discrepencies observed in the flavor sector of particle physics.  Recently many 
experiments have reported observed deviations from SM predictions in few observables 
such as $R(D^{*})$, $R_{K^{(*)}}$, muon $(g-2)$, etc. with statistical significance in the 
range of $\sim (2-4)\,\sigma$, which could be, if not due to statistical fluctuations, strong 
hints of NP. The focus will therefore be on 
the upcoming precision machines such as {\tt Belle-II} and {\tt LHCb}.

In this work we carry out a phenomenological study by extending the SM with new particles 
and show how this extension can explain the deviations in $R(D^{(*)})$ without violating any 
experimental constraints. To do this we add an additional color-singlet matter multiplet in 
the form of a vector-like lepton doublet under $SU(2)_L$. We also add a neutral scalar 
as well as an $SU(3)_{c}$ triplet scalar leptoquark (LQ), both singlets under the $SU(2)_L$ gauge group. The only additional 
requirement on all the additional particles is that they are odd under a discrete 
$\mathcal{Z}_{2}$ symmetry.  We then proceed to calculate the contributions from such a model to the $R(D^{(*)})$ and compare it to the observed deviations. We take into account all stringent constraints coming from flavor precision 
data on the model parameters including those coming from $K^{0}-\bar{K^{0}}$ and 
$B^{0}_{i}-\bar{B^{0}_{i}}$ ($i = d, s$) oscillations, $Br(Z \rightarrow f\bar{f})$ 
($f = u, d, s, b, e, \mu, \tau$) and from Peskin-Takeuchi S, T and U parameters. 
We show that the new particles can contribute substantially to $R(D^{*})$ provided 
the parameters are tuned in a phenomenological way.

Our paper is organised as follows. In section \ref{mod-det} we give the new particle 
content and their interaction Lagrangian. In section \ref{sect:constraints} we discuss the 
relevant constraints that would restrict out fit on the parameters of the model with 
respect to limits coming from $b \rightarrow c \tau \nu_{\tau}$, neutral meson oscillation 
data and $Z$-pole data. Finally in section \ref{sect:conclusions} we summarize our 
results and conclude.

\section{Model details.}
\label{mod-det} 
 
In this work we study a model which extends the SM particle content with a 
vector-like lepton $L_{4}$ = ($F^{0}$\ $F^{-})^{T}$, doublet under the $SU(2)_{L}$ and odd 
under a discrete $\mathcal{Z}_{2}$ symmetry, an $SU(3)_{c}$ triplet scalar leptoquark ($\phi_{LQ}$) odd under 
the $\mathcal{Z}_{2}$ and singlet under the $SU(2)_{L}$ gauge group carrying +$\frac{2}{3}$ unit of 
electic charge and a neutral complex scalar ($S$) singlet under the SM gauge group and 
odd under the $\mathcal{Z}_{2}$.  The quantum numbers under the SM gauge symmetry 
and the new discrete $\mathcal{Z}_{2}$ for the new particles are shown in Table \ref{tab1}.
Note that all the SM particles are even under the $\mathcal{Z}_{2}$. We write the most general Yukawa interaction Lagrangian involving the new set of particles that is consistent with all the symmetries of the model as
\begin{align}
\mathcal{L}_{NP} = \sum^{3}_{i=1}h_{i}\bar{Q_{i}}_{L} \, L_{4R}\phi_{LQ} + \sum^{3}_{j=1}h_{j}\bar{L_{j}}_{L} \, L_{4R}S + m_{F}\bar{L}_{4L}L_{4R} + h.c. %+ V(H,\phi_{LQ},S)
\label{Eq:Lep1}
\end{align}
where $i=1,2,3$ represent the SM quark generations and the couplings can be put as 
($h_u,h_c,h_t$) or equivalently ($h_d,h_s,h_b$). Similarly 
$j=1,2,3$ represent the SM lepton generations and the couplings can be put as ($h_e,h_\mu,h_\tau$) for the leptons while $m_{F}$ is the mass of 
the new vector-like leptons. 
%with $V(H,\phi_{LQ},S)$ denoting the scalar potential. The new particles and their charges under the symmetry groups are tabulated in Table \ref{tab1}.
\begin{table}[h!]
\begin{center}
\begin{tabular}[b]{|c|c|c|c|c|} \hline
Particles & $SU(3)_{c}$ & $SU(2)_{L}$ & $U(1)_{Y}$ & $\mathcal{Z}_{2}$ \\
\hline\hline
$L_{4}$ & 1 & 2 & -1/2  & -1 \\
\hline
$\phi_{LQ}$ & 3 & 1 & +2/3 & -1 \\
\hline
S & 1 & 1 & 0 & -1 \\
\hline
\end{tabular}
\end{center}
\caption{The charge assignments of new particles under the SM gauge group and 
$\mathcal{Z}_{2}$.}
\label{tab1}
\end{table}
With the additional scalar LQ $\phi_{LQ}$ and the complex scalar $S$, the most general scalar potential that is invariant under the full symmetry of the model can be written as \cite{Chaing-Wei-Chang}
\begin{align}
V(H,\phi_{LQ}, S) =& \, m^{2}H^{\dagger}H + m_{\phi_{LQ}}^{2}\phi^{\dagger}_{LQ}\phi_{LQ} + m_{S}^{2}S^{\dagger}S + \frac{\lambda_{1}}{4}(H^{\dagger}H)^{2} + \lambda_{HLQ}(H^{\dagger}H)(\phi_{LQ}^{\dagger}\phi_{LQ}) \nonumber \\
+& \lambda_{\phi S}(\phi_{LQ}^{\dagger}\phi_{LQ})(S^{\dagger}S) + \lambda_{HS}(H^{\dagger}H)(S^{\dagger}S) + \frac{\lambda_{\phi_{LQ}}}{4}(\phi_{LQ}^{\dagger}\phi_{LQ})^{2} + \frac{\lambda_{S}}{4}(S^{\dagger}S)^{2} \nonumber \\
+&  \left(\frac{m_{S_{1}}}{2}S^{2} + \frac{\lambda_{S_{1}}}{4}S^{4} + \frac{\lambda_{S_{2}}}{3}|S|^{2}S^{2} + \frac{\lambda_{HS}^{'}}{2}|H|^{2}S^{2} + h.c. \right) \nonumber \\ 
+&  \left( \frac{m_{\phi_{LQ1}}}{2}\phi_{LQ}^{2} + \frac{\lambda_{\phi_{LQ1}}}{4}\phi_{LQ}^{4} + \frac{\lambda_{\phi_{LQ2}}}{3}|\phi_{LQ}|^{2}\phi_{LQ}^{2} + \frac{\lambda_{H\phi_{LQ}}^{'}}{2}|H|^{2}\phi_{LQ}^{2} + h.c. \right)
\label{scalar-pot}
\end{align}
where $H$ represents the SM Higgs doublet. The new scalar fields do not get any vacuum 
expectation value (VEV) and can be expressed as
\begin{align}
\phi_{LQ} = \frac{\phi_{R} + i\phi_{I}}{\sqrt{2}}, &&   S = \frac{S_{R} + iS_{I}}{\sqrt{2}}.
\end{align}
Then we have a mass relation for the real and imaginary components of the scalars given 
by $m_{S_{R}} - m_{S_{I}} = m_{S_{1}} + \lambda_{HS}^{'}v_{0}$ where $v_{0}$ is the 
electroweak VEV for the SM Higgs. Note that for $m_{S_{R}} - m_{S_{I}} > 0$ the 
$S_{R}$ becomes the lightest component of the neutral singlet scalar $S$. 
As the $\mathcal{Z}_{2}$ remains unbroken, with $m_F$ larger than $m_{S_{R}}$ this
will be stable and can be a DM candidate.  However we find that to fit our 
results for $R(D^{*})$ we require that its Yukawa couplings have to be large with the  
fermions as suggested by $b \rightarrow c \tau \nu_{\tau}$ data. This would lead to large 
annihilation cross section and therefore its contribution to the present relic density is
 expected to be small \cite{Chaing-Wei-Chang} which is acceptable and not ruled out. 
 In this analysis, for simplicity we take $m_{S_{1}} = \lambda_{HS}^{'} \approx 0$ and 
 $m_{\phi_{LQ1}} = \lambda_{H\phi_{LQ}}^{'} \approx 0$ which means that for both the 
 new scalars their real part and complex part are symmetric in all respect.

\section{Constraints from neutral meson oscillation data, $Z$-pole and $b \rightarrow c \tau \nu_{\tau}$.}
\label{sect:constraints}

%In the SM the authors had not been able to construct an observable that is sensitive to quadrant in which the angles of CKM matrix belongs, but here we will show that 
We know that the Cabibbo-Kobayashi-Maskawa (CKM) mixing matrix with three real and 
one imaginary physical parameters can be made manifest by choosing an explicit 
parametrization. With the standard parameterization \cite{PDG2016} in terms if $
\theta_{ij}'s$ and the Kobayashi-Maskawa phase $\delta$ we find it interesting and worth
pointing out that the requirement of a positive contributions from NP to $R(D^{*})$ and 
constraints from neutral meson oscillations are favored when 
$\pi \leq \theta_{12} \leq \frac{3\pi}{2}$ and $\frac{3\pi}{2} \leq \theta_{13}, \theta_{23} \leq 
2\pi$. This implies that the sign of the first two rows of the CKM matrix elements are 
negative relative to the third row compared to instead the usual convention where all 
angles have been fixed in the first quadrant. When expressed in terms of the 
mass eigenstates of the down quarks ($d^{\prime}, s^{\prime}, b^{\prime}$), we have
\begin{align}
h_{i}^{\prime} = \sum^{j=t}_{j=u}h_{j}V_{ji}
\label{Eq:rotated}
\end{align}
for the down quark Yukawa couplings, where $i = d, s, b$.\footnote{The notation we use 
on the right side of Eq. (\ref{Eq:rotated}) is by representing $h_j$ as $h_u,h_c,h_t$ to 
write it in a compact way. However these are the same as $h_d,h_s,h_b$ respectively, as pointed out below Eq. (\ref{Eq:Lep1}) and as written explicitly in Eq. (\ref{Eq:cond1}).} Thus
we note that the effective coupling of the down-type quarks with the new vector-like leptons 
and the scalar LQ are modified in the mass basis via the CKM mixing matrix while the 
up-type quark couplings remain the same. Now we would like to point out that if we impose 
the condition
\begin{align}
h_{d}^{\prime} = -h_{d}V_{ud} - h_{s}V_{cd} + h_{b}V_{td} = 0,
\label{Eq:cond1}
\end{align}
then the NP has no contribution to the $K^{0}-\bar{K}^{0}$ and $B^{0}-\bar{B}^{0}$ 
oscillations. Since these observables are very precisely measured and no deviations from 
the SM prediction have been reported, the above condition seems a quite natural 
experimental imposition. Note that in Eq. (\ref{Eq:cond1}) the sign change of the first 
two rows of the CKM matrix elements is explicitly shown. In addition to this it is favorable to 
have significantly large Yuakawa coupling strength for the third generation interaction and 
we therefore choose the perturbative upper limit for the coupling 
$h_{b} = 3.52 < 2\sqrt{\pi}$ which favors the $b \rightarrow c \tau \nu_{\tau}$ data 
and parameterize $h_{s} \approx \frac{h_{b}}{a}$. Now if we demand $a$ to be 
real then to satisfy Eq. (\ref{Eq:cond1}), $h_{d}$ has to be complex. Additional 
constraint on the parameters also come from the respective mass bounds on the 
new charged and neutral leptons as well as bounds from $B^{0}_{s}-\bar{B}^{0}_{s}$ 
oscillation on $Re(\Delta M_{B^{0}_{s}}^{NP})$ and $Im(\Delta M_{B^{0}_{s}}^{NP})$. This 
is discussed in more detail in section \ref{subsect:Osc}.

\subsection{Contribution to $b \rightarrow c \tau \nu_{\tau}$.}
\label{subsect:bctnu}

In recent works in \cite{My-paper2-2017,My-paper1-2017}, it was shown that 
observed deviations from SM in the muon ($g-2$), generation of small neutrino masses, 
Baryogenesis as well as the observed anomalies in $R_{K^{(*)}}$ could be explained 
with new exotic scalars and leptons. Therefore it is very interesting to see whether 
exotic scalars and leptons can also explain the $R(D^{*})$, where 
$R(D^{*}) = \frac{Br(B \to D^{(*)} \tau \nu_{\tau})} {Br(B \to D^{(*)} \ell \nu_{\ell})}$ with 
$\ell = e, \mu$. A deviation from SM predictions in $R(D^{(*)})$ was first reported by 
{\tt Babar} \cite{RSinha3} followed by {\tt Belle} \cite{RSinha4,RSinha5,RSinha6} and 
{\tt LHCb} \cite{RSinha7,RSinha8}, with the latest HFAG average of the 
experimental result amounting to \cite{RSinha9}
\begin{align}
R(D)^{Exp} = 0.407 \pm 0.039 \pm 0.024 \,\, ;
&&
R(D^{*})^{Exp} = 0.304 \pm 0.013 \pm 0.007.
\end{align}
These when compared to the SM predictions as given in \cite{RSinha15,RSinha16} respectively:
\begin{align}
R(D)^{SM} = 0.300 \pm 0.008 \,\,;
&&
R(D^{*})^{SM} = 0.252 \pm 0.003,
\end{align}
and taking the correlation between the two observables into account, the combined 
deviation from SM  is around 4.1$\sigma$ in these observables. Although the present 
HFAG world averages are well above the SM predicted values, the {\tt Belle} results agrees 
with both the SM value as well as the HFAG world averages \cite{RSinha2017}, where 
HFAG world averages in these observables are still dominated by the {\tt Babar}'s data due 
to it having the least error of all the measurements till date.

\begin{figure}[h!]
\hspace{0.4cm}
\includegraphics[width=3.6in, height=1.8in]{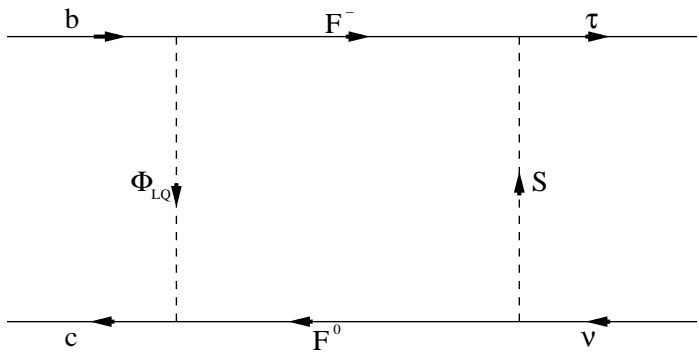}
\caption{Contributions to the $b \rightarrow c \tau \nu_{\tau}$ from the new particles at box loop level.}
\label{Fig1:fig2}
\end{figure}

In this model, there is no contribution to $b \rightarrow c \tau \nu_{\tau}$ transition at tree 
level, but at the box loop level there is contribution from NP to the quark level transition due 
to the Yukawa interactions shown in Eq. (\ref{Eq:Lep1}). The box loop diagram shown in 
Figure \ref{Fig1:fig2} from NP add coherently to the SM contribution and so we can express 
the effective Hamiltonian as \cite{RSinha2017}
\begin{align}
\mathcal{H}^{eff} = \frac{4G_{F}}{\sqrt{2}}V_{cb}(1 + C^{NP})[(c,b)(\tau , \nu_{\tau})]
\end{align}
where $(c,b)(\tau , \nu_{\tau})$ is the usual SM left handed vector four current operator and $C^{NP}$ in our model can be expressed as
\begin{align}
C^{NP} = \mathcal{N}\frac{(-V_{ub}h_{d} - V_{cb}h_{s} + V_{tb}h_{b})^{*}|h_{s}||h_{\tau}|^{2}}{64\pi^{2}m_{F}^{2}}S(x_{i},x_{j}),
\end{align}
where 
\begin{align*}
S(x_{i},x_{j}) = \frac{1}{(1 - x_{i})(1 - x_{j})} + \frac{x_{i}^{2}\ln(x_{i})}{(1 - x_{i})^{2}(x_{i} - x_{j})} - \frac{x_{j}^{2}\ln(x_{j})}{(1 - x_{j})^{2}(x_{i} - x_{j})} 
\end{align*}
are the Inami-Lim functions \cite{LDLuzio62,ACrivellin} with 
\begin{align*}
\frac{1}{\mathcal{N}} = \frac{4G_{F}|V_{cb}|}{\sqrt{2}}, && x_{i} = \frac{m^{2}_{\phi_{LQ}}}{m^{2}_{F^{-}}} && {\rm and} && x_{j} = \frac{m^{2}_{S}}{m^{2}_{F^{0}}}.
\end{align*}

%We have evaluated the box loop using the formulas given in the Appendix C of \cite{FranzGross-book}. 
%and setting all the external momenta to zero compared to the 
%masses of the particles in the loop \cite{Cheng-Li-book}. 
We have cross checked our calculations and find it to agree with a similar 
result evaluated in context of $b \rightarrow s \mu^{+}\mu^{-}$ given in \cite{ACrivellin} and we taken 
$\lambda = 0.22506 \pm 0.00050$, $A = 0.811 \pm 0.026$, $\bar{\eta} = 0.356 \pm 0.011$ 
and $\bar{\rho} = 0.124^{+0.019}_{-0.018}$, ignoring corrections of 
$\mathcal{O}(\lambda^{4})$ and above, where $\lambda$, $A$, $\bar{\eta}$ and 
$\bar{\rho}$ being the parameters in the Wolfenstein parametrization of the CKM matrix 
elements \cite{PDG2016}. 

We choose $h_{e}, h_{\mu} << h_{\tau} = 3.52$  along with fixing the benchmark value 
of the masses well above the present respective experimental bounds \cite{PDG2016}.  
 For our calculation we assume $m_{F^{\pm}} = m_{F^{0}} = 200$ GeV, 
 $m_{\phi_{LQ}} = 900$ GeV and $m_{S} = 150$ GeV. Further we choose $a$ 
 ($a \approx \frac{h_{b}}{h_s}$) to be real along with the constraints on Yukawa 
 couplings from Eq. (\ref{Eq:cond1}) as well as requiring 
 $Re[(h_{s}^{'}h_{b}^{'})^{2}] \leq 3.197\times 10^{-3}$ from $\Delta M_{SM}^{error}$ and 
 $Im[(h_{s}^{'}h_{b}^{'})^{2}] \leq 2.617\times 10^{-3}$ from CP violation data in 
 $B^{0}_{s}-\bar{B}^{0}_{s}$ oscillation (see section \ref{subsect:Osc} 
 for details). We get for the best fit values of the parameters as $a = -21.588$, 
 $Re(h_{d}) = -8.402\times 10^{-3}$ and $Im(h_{d}) = -0.0119$ which gives
\begin{align}
R(D^{*})^{NP} = 0.263 \pm 0.051
&&
{\rm and}
&&
R(D)^{NP} = 0.313 \pm  0.158.
\end{align}
Compared to the experimental values there is substantial contribution especially to the 
$R(D^{*})$ from the NP, where the errors quoted here are the experimental errors scaled by 
$\sqrt{\chi^{2}}.$\footnote{where $\chi^{2} = [\frac{(R(D)^{Exp}-R(D)^{NP})^{2}}{\sigma_{Exp}^{2}(D)} + \frac{(R(D^{*})^{Exp}-R(D^{*})^{NP})^{2}}{\sigma^{2}_{Exp}(D^{*})}]$.} 
\begin{figure}[hb!]
\includegraphics[height=2.4in,width=3.2in]{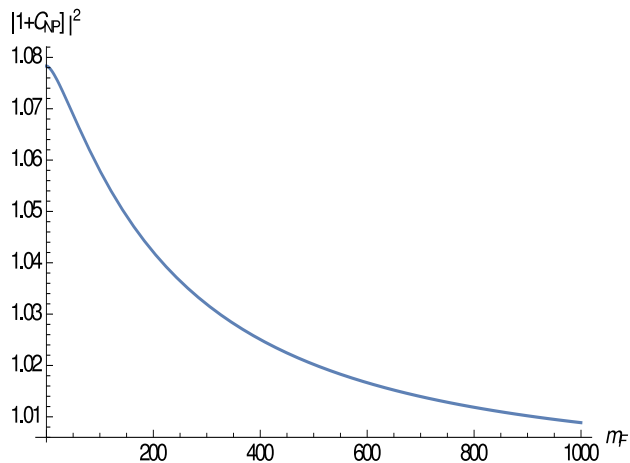}
\caption{Plot of $|1 + C_{NP}|^{2}$ vs $m_{F}$ for $m_{LQ} = 900$ GeV and $m_{S} = 150$ GeV.}
\label{Fig1:fig11}
\end{figure}
The contribution from NP has reduced the deviation in $R(D^{*})$ from $3.4\sigma$ to 
$0.8\sigma$ and deviation in $R(D)$ from 2.3$\sigma$ to 0.6$\sigma$. In 
Figure \ref{Fig1:fig11} we plot $|1 + C_{NP}|^{2}$ as a function of $m_{F}$ while 
fixing $m_{LQ} = 900$ GeV and $m_{S} = 150$ GeV. The NP contribution goes down as 
$m_F$ increases as shown in Figure \ref{Fig1:fig11} and falls to 1\% of the SM value as 
$m_{F} \sim 1$ TeV.

The $(c,b)(\tau,\nu_{\tau})$ current can also contribute to $B_{c} \rightarrow \tau \nu_{\tau}$ 
with same $|1+C^{NP}|^{2} = 1.04208$, however this is much smaller than the present 
allowed limit given as $|1+C^{NP}|_{allowed}^{2} = 1.69$ \cite{RAlonso}. Similarly NP 
contribution to $D_{s} \rightarrow \tau \nu_{\tau}$ is negligible compared to SM as well.  
Note that with the additional particle content of the model and their Yukawa interaction 
terms we also get NP contributions to hadronic decay of $\tau$.
The decay $\tau \rightarrow (K\pi)\nu_{\tau}$ is proportional to the product of couplings 
$h_{s}^{\prime}h_{d}|h_{\tau}|^{2}$ and gives $C_{NP}^{\prime} = \mathcal{O}(10^{-6})$. 
Thus the NP contribution is quite small and negligible compared to SM contributions in 
this mode. Even though $h_{d}$ is complex, it still cannot contribute to the CP violation 
in $\tau \rightarrow (K\pi)\nu_{\tau}$ or $\tau \rightarrow \rho\pi\nu_{\tau}$ etc. This is 
because the NP contributions in this model to the vector and the axial-vector effective 
four current come with same magnitude and phase (see Ref. \cite{MyCP-paper, MyCP-conf-paper} and references there in for more details). 
NP contributions to $C_{9}$ in $B \rightarrow K^{(*)}\mu^{+}\mu^{-}$ via photon penguin 
is about $|C^{NP}_{9}| = 3.428\times 10^{-3}$ which is again too small to have any effect 
on the reported anomaly in $C_{9}$ \cite{ACrivellin} and NP contributions to 
$b \rightarrow s \gamma$ is $|C^{NP}_{7} + 0.24C^{NP}_{8}| \approx 10^{-3}$ which is 
about two orders smaller than the 2$\sigma$ present experimental bound \cite{ACrivellin}. 
In this model we can also get contributions to $B \rightarrow K^{(*)}\tau^{+}\tau^{-}$, 
$B_{s} \rightarrow \tau^{+}\tau^{-}$ and $D^{0} \rightarrow (\pi^{0})\nu_{\tau}\bar{\nu}_{\tau}$ 
which are not properly measured yet but NP contributions to these modes are less than a 
percent-level, at the order of $|1+C^{NP}_{2}|^{2} = 1.0042$ or smaller and so negligible 
compared to the SM contribution. We also note that NP contribution to the 
anomalous magnetic moment of $\tau$ is $\Delta a^{NP}_{\tau} \approx -3.9\times 10^{-8}$ 
compared to the experimental bound 
$-0.052 < \Delta a^{Exp.}_{\tau} < 0.013$ \cite{PDG2016} which is again negligible.

\subsection{Neutral meson oscillation.}
\label{subsect:Osc}

\begin{figure}[h!]
\includegraphics[width=6.0in, height=2.0in]{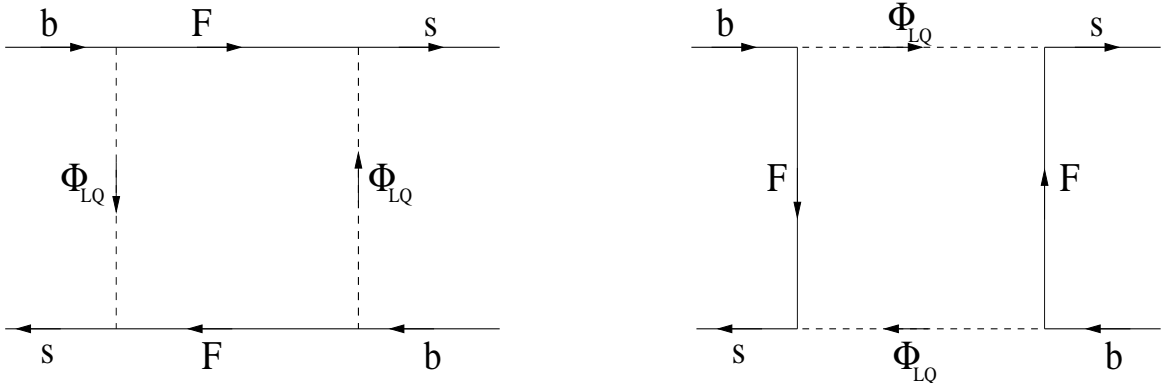}
\caption{Contributions to the $B^{0}_{s}-\bar{B}^{0}_{s}$ mixing from the new particles. The
$F$ in the loop is the charged component of the vector-like lepton doublet.}
\label{Fig1:fig1}
\end{figure}

Similar to the SM, the new particles in our model also contribute to neutral 
meson oscillations via the box loop. From the condition that we imposed in 
Eq.(\ref{Eq:cond1}) our model gives no contribution to the 
$K^{0}-\bar{K}^{0}$ and $B^{0}-\bar{B}^{0}$. However for 
$B^{0}_{s}-\bar{B}^{0}_{s}$ oscillations we do have non vanishing contributions 
which can be put as
\begin{align}
\mathcal{L}^{NP}_{eff} = 2 C_{B_{s}}^{NP}\bar{s_{\alpha}}\gamma^{\mu}P_{L}b_{\alpha}\bar{b_{\beta}}\gamma_{\mu}P_{L}s_{\beta}
\end{align}
where $C_{B_{s}}^{NP} = \frac{(h^{'}_{s}h^{'}_{b})^{2}}{128\pi^{2}m_{F}^{2}}S(x,x)$ 
where $S(x,x)$ are again the Inami-Lim functions 
with $x = \frac{m_{LQ}^{2}}{m_{F}^{2}}$ and the factor 2 to account for the contributions from the two diagrams in Figure \ref{Fig1:fig1}. 
Then introducing a factor of $\frac{1}{2!2!}$ to take into account the Wick contraction and color structure over 
counting\footnote{When expanded in terms of creation and annihilation operators, there will be four terms, two from each diagrams, only two will contribute to the real process but when summed all four are summed so a factor $\frac{1}{2!}$ to compensate it; and also when summed over the colors, we sum over the two different possible color singlet arrangements but only one actually contribute, so a factor $\frac{1}{2!}$ to compensate that, actually these over counted factor of 4 are in $\frac{8}{3}$ factor in the Eqs.(\ref{Eqs:Del-M}), see \cite{Cheng-Li-book, Leader-Predazzi-book} for details.}, we 
 have
\be
\langle B^{0}_{s}|\bar{s_{\alpha}}\gamma^{\mu}P_{L}b_{\alpha}\bar{b_{\beta}}\gamma_{\mu}P_{L}s_{\beta}|\bar{B}^{0}_{s} \rangle = \frac{1}{4}\times\frac{1}{4}\times\frac{8}{3}M^{2}_{B^{0}_{s}}f^{2}_{B_{s}^{0}}B(\mu)\frac{1}{2M_{B^{0}_{s}}}
\ee
and so with $\Delta M_{B^{0}_{s}}^{NP} = 2Re(\langle B^{0}_{s} |\mathcal{L}^{NP}_{eff}| \bar{B}^{0}_{s} \rangle$) we have
\be
\Delta M_{B^{0}_{s}}^{NP} =  \frac{1}{4\times 2}\times \frac{8}{3}M_{B^{0}_{s}}f^{2}_{B_{s}^{0}}B(\mu)\times C_{B_{s}}^{NP},
\label{Eqs:Del-M}
\ee
where $f_{B^{0}_{s}}$ is the $B^{0}_{s}$ decay form factor and $B(\mu)$ is a QCD scale correction factor, their values are taken from \cite{LDLuzio60, LDLuzio2018}.

Then with the values of the Yukawa couplings and masses given in the previous section we 
get $Re(\Delta M^{NP}_{B^{0}_{s}}) = 1.243$ ps$^{-1}$. This when compared to the 
error in experimental measurement of the same observable given as 
$\Delta M^{Exp}_{B^{0}_{s}} = (17.757 \pm 0.021)$ ps$^{-1}$, the NP contribution is well 
above the error in the experimental measurement taken from PDG \cite{PDG2016}. 
But given that there are still large errors in the SM calculations with the latest estimate of 
SM calculation predicting a 1.8$\sigma$ deviation above the experimental average as 
$\Delta M^{SM}_{B^{0}_{s}} = (20.01 \pm 1.25)$ ps$^{-1}$ \cite{LDLuzio2018}, the above 
NP contribution is well within the error in the latest SM calculation. We would like to point 
out that the previous SM calculations in Refs. \cite{LDLuzio60,LDLuzio61} agrees 
with the experimental value but their errors are larger than the latest SM prediction. 
Since NP contribution is allowed to be as large 
as the SM and experimental errors added in quadrature, if we take the previous SM 
predictions, NP contribution is allowed to be little larger than the above value. Note that in 
all the above calculations we have taken the hadronization parameters 
from Refs. \cite{LDLuzio60, LDLuzio2018} and the experimental values from 
PDG \cite{PDG2016}. 

Due to $h_{d}$ being complex, we also have a non-zero imaginary 
component of $\Delta M^{NP}_{B^{0}_{s}}$ given as 
$Im(\Delta M^{NP}_{B^{0}_{s}}) = -0.715\times \Gamma^{Exp}_{B^{0}_{s}}$. This can 
contribute to the CP violation in the $B^{0}_{s}-\bar{B}^{0}_{s}$ mixing which is 
parametrized in terms of $\frac{Re(\epsilon_{B^{0}_{s}})}{1 + |\epsilon_{B^{0}_{s}}|^{2}}$, 
where $\epsilon_{B^{0}_{s}} = \frac{-\frac{1}{2}Im(\Delta M_{B^{0}_{s}})}{\frac{1}{2}\Delta \Gamma_{B^{0}_{s}} - i\Delta M_{B^{0}_{s}}}$.
%, see page 374 of \cite{Cheng-Li-book} for a derivation in a general oscillation context. 
In our case with $\Delta \Gamma_{B^{0}_{s}} << \Delta M_{B^{0}_{s}}$ the CP violating 
parameter due to NP can be approximately expressed as 
$\frac{Re(\epsilon^{NP}_{B_{s}^{0}})}{1 + 
|\epsilon^{NP}_{B_{s}^{0}}|^{2}} \approx \frac{-Im(\Delta M^{NP}_{B_{s}^{0}})\times 
\Delta \Gamma^{Exp}_{B^{0}_{s}}}{4(\Delta M_{B^{0}_{s}}^{Exp})^{2}} \approx 
+1.050\times 10^{-5}$ compared to 
$\frac{Re(\epsilon^{Exp})}{1 + 
|\epsilon^{Exp}|^{2}} \approx (-1.5 \pm 7)\times 10^{-4}$ \cite{PDG2016}. Note that the 
NP contribution is an order of magnitude smaller than the present experimental limit. 
There is no contribution to the $\Delta \Gamma_{B^{0}_{s}}$ from NP since none of 
the intermediate particles in Figure \ref{Fig1:fig1} can go on shell. For the 
$D^{0}-\bar{D}^{0}$ oscillation with 2$\sigma$ bound from \cite{ACrivellin} given as 
$|C^{Exp}_{D^{0}}| < 2.7\times 10^{-7}$ $TeV^{-2}$ we compare 
$|C^{NP}_{D^{0}}| \approx 2.971\times 10^{-9}$ $TeV^{-2}$ and  
find the NP contribution to be around two orders of magnitude smaller than 
the present experimental bound at 2$\sigma$.

\subsection{Z pole constrains.}
\label{Z-pole}
For theoretical calculations of contribution from new fermions to the $Z$ decay into 
two fermions via higher order loops, we have used \cite{Chaing-Wei-Chang}
\begin{align}
Br(Z \rightarrow f_{i}f_{i}) = \frac{G_{F}}{3\sqrt{2}\pi}\frac{m_{Z}^{3}}{(16\pi^{2})^{2}\Gamma^{tot.}_{Z}}(T^{i}_{3} - Q^{i}\sin^{2}(\theta_{W}))^{2}|h_{i}^{'}|^{4}|[F_{2}(m_{F},m_{\phi}) + F_{3}(m_{F},m_{\phi})]|^{2}
\end{align}
where
\begin{align}
F_{2}(a, b) = \int^{1}_{0}dx(1-x)\ln{[(1-x)a^{2} + xb^{2}]}
\end{align}
and
\begin{align}
F_{3}(a, b) =& \int^{1}_{0}dx\int^{1-x}_{0}dy\frac{(xy-1)m_{Z}^{2} + (a^{2}-b^{2})(1-x-y) - \Delta\ln\Delta}{\Delta} \\
\Delta =& -xym^{2}_{Z} + (x+y)(a^{2}-b^{2}) + b^{2}
\end{align}
with $\Gamma^{tot}_{Z} = 2.4952$.

Now with the numerical values of the Yukawa couplings given before and  with
$m_{F^{\pm}} = m_{F^{0}} = 200$ GeV, $m_{\phi_{LQ}} = 900$ GeV and 
$m_{S} = 150$ GeV, we get $Br(Z \rightarrow \bar{d}d)_{NP} =0$ due to 
Eq. (\ref{Eq:cond1}) while $Br(Z \rightarrow \bar{u}u)_{NP}$, 
$Br(Z \rightarrow \bar{s}s)_{NP} \,<< \, 
Br(Z \rightarrow \bar{c}c)_{NP} \approx \mathcal{O}(10^{-10})$ well within the 
experimental errors given by $Br(Z \rightarrow \bar{u}u)^{Exp}_{error} \approx 0.004$, 
$Br(Z \rightarrow \bar{s}s)^{Exp}_{error} \approx 0.004 $ and 
$Br(Z \rightarrow \bar{c}c)^{Exp}_{error} \approx 0.0021$.  Even for the decay mode where 
the large Yukawa choices can be significant we find 
$Br(Z \rightarrow \bar{b}b)_{NP} = 4.737\times 10^{-5}$ as compared to 
$Br(Z \rightarrow \bar{b}b)^{Exp}_{error} \approx 5\times 10^{-4}$ putting the NP 
contribution an order of magnitude smaller than the experimental error. The contributions 
from NP to $Br(Z \rightarrow \bar{e}e)$ and $Br(Z \rightarrow \bar{\mu}\mu)$ are 
negligible compared to the experimental errors since we assume that $h_{e},\ h_{\mu} << 1$
(which is required to explain the $R(D^{(*)})$ anomalies).  For the third generation 
lepton where we have $h_{\tau}$ large we get 
$Br(Z \rightarrow \bar{\tau}\tau)_{NP} \approx 7.62\times 10^{-9}$  and 
$Br(Z \rightarrow \bar{\nu}\nu(invisible))_{NP} \approx 2.47\times 10^{-8}$ compare to 
$Br(Z \rightarrow \bar{\tau}\tau)_{error}^{Exp} \approx 8\times 10^{-5} < Br(Z \rightarrow 
invisible)_{error}^{Exp}$. Here again the NP contributions are negligible. All the 
experimental values are taken from the latest PDG averages \cite{PDG2016}. 
Regarding the contributions of the new states \cite{Hong-Jian-He6}, to the 
Peskin-Tekeuchi S, T and U parameters, we find that with the 
above given masses of the new fermions we have $S \approx 0.0203$, $T \approx 0$ and $U \approx 0$ in our model, which are well within the present experimental bounds on 
these parameters \cite{Hong-Jian-He-2001}.

\section{Conclusions.}
\label{sect:conclusions}

In this work we have introduced a vector like fourth generation lepton doublet 
($F^{0},\ F^{-}$) along with an $SU(3)_{c}$ triplet scalar leptoquark $\phi_{LQ}$ and a neutral scalar ($S$) 
both singlet under the $SU(2)_{L}$ gauge group. All the newly added particles are odd under a 
discrete symmetry $\mathcal{Z}_{2}$. With these new particles we have done a 
comprehensive analysis of the phenomenological consequences of the model taking all 
the very stringent constraints from $K^{0}-\bar{K^{0}}$ and 
$B^{0}_{i}-\bar{B^{0}_{i}}$ ($i = d, s$) oscillations as well as 
$Br(Z \rightarrow f\bar{f})$ and Peskin-Tekuchi parameters into account. We find that such 
a model can give a substantial contribution to $R(D^{(*)})$, and is able to reduce the 
tension between theoretical prediction and experimental measured value of $R(D^{*})$ 
from $3.4\sigma$ to $0.8\sigma$ and deviation in $R(D)$ from $2.3\sigma$ to 
$0.6\sigma$. Especially the NP contribution is able to reduce the discrepancy 
between experiment and theory in $R(D^{*})$ substantially. In addition the 
mass of the newly introduced states required to give a large contribution to 
the $R(D^{*})$ lie in a range which will be directly probed at the LHC with higher luminosity. 
Thus the model presents robust phenomenological consequences accessible at both 
the high energy collider experiment such as the LHC as well as leaving imprints in the 
flavor sector. 

We find that while accommodating the large contributions to $R(D^{(*)})$ the model does not violate any other observations and is found to satisfy all other stringent constraints coming 
from neutral meson oscillations and precision Z-pole data.

\acknowledgments{LD would like to thank Biswarup Mukhopadhyaya for 
very helpful discussions. LD would also like to thank Zoltan Ligeti for replying to his 
query regarding the observability of relative sign between the rows of CKM matrix. 
This work was partially supported by funding available from the Department of Atomic Energy, Government of India, for the Regional Centre for Accelerator-based 
Particle Physics (RECAPP), Harish-Chandra Research Institute.}

\end{document}